\documentclass[]{revtex4}
\usepackage{graphicx}

\begin{document}
\title{ Out-of-equilibrium chiral and $U_A(1)$ symmetry breaking in electromagnetic fields}
\author{Xingyu Guo and Pengfei Zhuang}
\address{Physics Department, Tsinghua University and Collaborative Innovation Center of Quantum Matter, Beijing 100084, China}
\date{\today}
\begin{abstract}
We systematically study quantum effect on chiral and $U_A(1)$ symmetry breaking under external electromagnetic fields in the frame of equal-time Wigner function formalism. We derive the transport and constraint equations for the quark distribution functions and the chiral and pion condensates in a Nambu--Jona-Lasinio model. By taking semi-classical expansion of the equations, chiral symmetry is broken at classical level, while $U_A(1)$ symmetry breaking happens only at quantum level. Beyond quasi-particle approximation, the quark off-shell effect leads to strong oscillation for the chiral and pion condensates.
\end{abstract}
\maketitle

In the chiral limit with zero current quark mass, the Lagrangian density of the quantum chromodynamics (QCD) respects the symmetry of $U_L(3)\times U_R(3) = U_V(1)\times U_A(1)\times SU_L(3)\times SU_R(3)$ at classical level. However, the quark-antiquark condensate $\langle\bar\psi\psi\rangle$ spontaneously breaks the $SU_L(3)\times SU_R(3)$ symmetry, and the $U_A(1)$ symmetry is broken by the quantum anomaly due to the nontrivial topology of the principal bundle of gauge field~\cite{thooft,leutwyler}. Chiral symmetry breaking leads to a rich meson and baryon spectrum, and as a supplement the $U_A(1)$ anomaly explains the nondegeneracy of $\eta$ and $\eta'$ mesons~\cite{witten,veneziano,rosenzweig}.

Like temperature and chemical potential, strong electromagnetic fields provide another way to change the QCD symmetries. For chiral symmetry, lattice simulations and effective model calculations show that, the magnetic field enhances the chiral symmetry breaking in vacuum which is called magnetic catalysis but reduces the critical temperature of the symmetry restoration which is called inverse magnetic catalysis~\cite{lattice1,lattice2,lattice3,gusynin,alexandre,ferrer,fraga,gatto,preis,fukushima,andersen,kamikado,ferreira,mao,evans,dudal}. As a gauge field, the parallel electromagnetic fields can also break down the $U_A(1)$ symmetry~\cite{cao,wang}, characterized by the pseudoscalar condensate $\langle\bar\psi i\gamma_5\tau_3\psi\rangle$ via the electromagnetic triangle anomaly process $\pi_0\to 2\gamma$.

In laboratories, the electromagnetic fields which are strong enough to modify the QCD properties can be formed only in the beginning of high energy nuclear collisions~\cite{kharzeev,skokov,deng}. Since the fields are created by the spectators of the collisions, they can be considered as external or classical fields for the parton motion in the fireball. Considering that the parton system in the initial stage is in nonequilibrium state, the interaction between the partons and the electromagnetic fields should be described in the frame of a kinetic theory. Recently, a quantum kinetic equation to describe the chiral fermion motion in electromagnetic fields is developed via different methods and applied to heavy ion collisions~\cite{stephanov,son,chen,huang,lucas}. In this paper, we study the out-of-equilibrium $U_A(1)$ symmetry breaking in electromagnetic fields and its relation to the chiral symmetry breaking in the frame of Wigner function formalism~\cite{degroot,heinz,elze,elze2}.

We effectively describe the quantum anomaly induced $U_A(1)$ symmetry breaking and spontaneous chiral symmetry breaking in a Nambu--Jona-Lasinio (NJL) model~\cite{njl} at quark level~\cite{vogl,klevansky,volkov,hatsuda,buballa} with scalar and pseudoscalar interaction channels. The Lagrangian density of the two-flavor model in electromagnetic fields is
\begin{equation}
\label{njl}
\mathcal L = \bar\psi\left(i\gamma^\mu D_\mu-m_0\right)\psi+G\sum_{a=0}^3\left[\left(\bar\psi\tau_a\psi\right)^2+\left(\bar\psi i\gamma_5\tau_a\psi\right)^2\right],
\end{equation}
where the covariant derivative $D_\mu=\partial_\mu+iQA_\mu$ couples the quark field $\psi$ with electric charge $Q = diag(Q_u, Q_d) = diag(2/3e, -1/3e)$ to the electromagnetic fields $A_\mu$, $G$ is the four-fermion coupling constant, and $\tau_a$ are the Pauli matrices in flavor space. In the chiral limit with vanishing current quark mass $m_0 = 0$ and vanishing electromagnetic fields, the Lagrangian density is with symmetry $U_V(1) \times U_A(1) \times SU_L(2) \times SU_R(2)$. For the chiral symmetry $SU_L(2) \times SU_R(2)$, it is explicitly broken down to $U_L(1) \times U_R(1)$ by the electromagnetic fields, and the rest of the symmetry $U_L(1) \times U_R(1)$ is then spontaneously broken by the chiral condensate $\sigma=2G\langle\bar\psi\psi\rangle$. For the $U_A(1)$ symmetry, it is spontaneously broken by the neutral pion condensate $\pi=2G\langle\bar\psi i\gamma_5\tau_3\psi\rangle$. The chiral and $U_A(1)$ symmetry breaking in electromagnetic fields in the NJL model was investigated in the case of thermal equilibrium~\cite{mao,cao}. We study here the out-of-equilibrium evolution of the symmetries. As the isospin-singlet neutral pseudo-scalar $\eta$ meson (more precisely, the two-flavor counterpart of the $\eta$ meson) has the same quantum numbers as $\pi_0$ meson except for isospin, one can expect that $\eta$ would also condensate via the electromagnetic chiral anomaly. Similarly, there should be not only the isospin-singlet condensate $\sigma$ but also the isospin-triplet condensate $a_3$, when the electromagnetic fields are turned on. In thermal equilibrium case, the properties of these condensates induced by electromagnetic chiral anomaly were investigated in the NJL model~\cite{wang}. Here we focus on the non-equilibrium neutral pion condensate and neglect the $\eta$ and $\eta'$ mesons.

The covariant quark Wigner function $W(x,p)$ is the ensemble average of the Wigner operator which is the four-dimensional Fourier transform of the covariant density matrix~\cite{schwinger},
\begin{equation}
\label{w4}
W(x,p) = \int d^4 y e^{ipy}\langle\psi(x_+)e^{iQ\int_{-{1\over 2}}^{1\over 2} dsA(x+sy)y}\bar\psi(x_-)\rangle
\end{equation}
with $x_\pm = x\pm y/2$, where the exponential function of the electromagnetic fields guarantees the gauge invariance of the Wigner function. Note that, when the gauge fields are external fields, the gauge link in the Wigner function can be taken out from the ensemble average. We restrict ourselves to the kinetic equations for the Wigner function in the mean field approximation, by replacing the field operators $\bar\psi\psi$ and $\bar\psi i\gamma_5\tau_a\psi$ by their mean values. Such a Hartree approximation has so far been used in most applications of quantum transport theory. Under the assumption of external electromagnetic fields, the Dirac equation for the quark field $\psi$ is written as,
\begin{equation}
\label{dirac}
\left[i\gamma^\mu D_\mu-(m_0-\sigma)+i\gamma_5\tau_3\pi\right]\psi(x)=0.
\end{equation}
Calculating the first-order derivatives of the covariant density matrix in the Wigner function with respect to $x$ and $y$ and using the Dirac equation, one obtains the covariant kinetic equation
\begin{equation}
\label{ck}
\left(\gamma^\mu K_\mu+\gamma_5\tau_3 K_5-M\right)W(x,p)=0
\end{equation}
with the electromagnetic, scalar and pseudoscalar operators
\begin{eqnarray}
\label{4operator}
K_\mu &=& \Pi_\mu +{i\hbar\over 2}D_\mu,\nonumber\\
K_5 &=& \Pi_5 +{i\hbar\over 2}D_5,\nonumber\\
M &=& M_1+iM_2,\nonumber\\
\Pi_\mu &=& p_\mu-iQ\hbar \int_{-1/2}^{1/2}ds s F_{\mu\nu}(x-i\hbar s\partial_p)\partial_p^\nu,\nonumber\\
D_\mu &=& \partial_\mu-Q\int_{-1/2}^{1/2}ds F_{\mu\nu}(x-i\hbar s \partial_p)\partial_p^\nu,\nonumber\\
\Pi_5 &=& \sin\left({\hbar\over 2}\partial_x\partial_p\right)\pi(x),\nonumber\\
D_5 &=& \cos\left({\hbar\over 2}\partial_x\partial_p\right)\pi(x),\nonumber\\
M_1 &=& m_0-\cos\left({\hbar\over 2}\partial_x\partial_p\right)\sigma(x),\nonumber\\
M_2 &=& \sin\left({\hbar\over 2}\partial_x\partial_p\right)\sigma(x)
\end{eqnarray}
with the electromagnetic tensor $F_{\mu\nu}(x)=\partial_\mu A_\nu(x) - \partial_\nu A_\mu(x)$. We have explicitly shown the $\hbar$-dependence in order to be able to discuss the semiclassical expansion of the kinetic equation in the following. While the kinetic equation (\ref{ck}) and the operators (\ref{4operator}) here are similar to what derived in Ref.\cite{zhuang}, $\sigma$ and $\pi$ are independent external parameters there like the external electromagnetic field $A_\mu$ but expressed in terms of the quark Wigner function here,
\begin{eqnarray}
\label{sigmapion}
\sigma(x) &=& G\int{d^4 p\over (2\pi)^4}\text {Tr}W(x,p),\nonumber\\
\pi(x) &=& G\int{d^4 p\over (2\pi)^4}\text{Tr}\left[i\gamma_5W(x,p)\right].
\end{eqnarray}
Therefore, from the solution of the kinetic equation we can self-consistently obtain the dynamical evolution of the two order parameters, see the following.

The Wigner function in spinor QED is a complex $4\times 4$ matrix in Dirac space. Vasak, Gyulassy and Elze discussed its spin decomposition and derived a set of coupled equations for the spin components~\cite{vasak}. Considering the flavor dependent interaction in the NJL model (\ref{njl}), the Wigner function $W(x,p)$ here is defined in Dirac space and flavor space, and we should do its spin and isospin decomposition,
\begin{equation}
\label{decom}
W = \sum_{q=u,d}{\tau_q\over 8}\left[F_q+i\gamma_5 P_q+\gamma_\mu V_q^\mu+\gamma_5\gamma_\mu A^\mu_q+{\sigma_{\mu\nu}\over 2}S^{\mu\nu}_q\right],
\end{equation}
where the matrices $\tau_q$ in flavor space are defined as $\tau_u = diag(1,0)$ and $\tau_d=diag(0,1)$. The components $\Gamma_{iq}=\{F_q, P_q, V^\mu_q, A^\mu_q, S^{\mu\nu}_q\}$ are real functions and can thus be interpreted as physical phase-space densities. Substituting the decomposition (\ref{decom}) into the kinetic equation (\ref{ck}), the equations for the components can be divided into two groups. One contains the constraint equations which are the extension of the classical on-shell condition, and the other contains the transport equations which are the extension of the classical Boltzmann equation~\cite{vasak,zhuang2}.

In comparison with the covariant formalism of a relativistic kinetic theory, the equal-time formalism~\cite{birula,zhuang,zhuang2,zhuang3,ochs} can be solved as a well defined initial value problem, especially when the particles are not on the shell. It is thus conveniently used in quantum many body systems. The equal-time Wigner function ${\cal W}(x,{\bf p})$ and its spin and isospin decomposition are defined as
\begin{eqnarray}
{\cal W} &=& \int dp_0 W\gamma_0\nonumber\\
&=& \sum_{q=u,d}{\tau_q\over 8}\big[f^q_V+\gamma_5 f_{PV}^q-i\gamma_0\gamma_5 f_P^q+\gamma_0 f_S^q
+\gamma_5\gamma_0{\bf \gamma}\cdot{\bf g}_{PV}^q+\gamma_0{\bf \gamma}\cdot{\bf g}_V^q-i{\bf \gamma}\cdot {\bf g}_{T0}^q-\gamma_5{\bf \gamma}\cdot {\bf g}_T^q\big],
\end{eqnarray}
where the components $f_S, f_P, f_V, {\bf g}_V, f_{PV}, {\bf g}_{PV}, {\bf g}_{T0}$ and ${\bf g}_T$ correspond respectively to the scalar, pseudoscalar, vector, pseudovector and tensor channels of the interaction. Performing an energy average of the two groups of covariant kinetic equations, we obtain the transport equations,
\begin{eqnarray}
\label{transport}
&& \hbar(D_q f^q_V+{\bf D}_q\cdot{\bf g}^q_V)+2m_2 f^q_S-2\pi^q_5 f^q_P=0,\nonumber\\
&& \hbar(D_q f^q_{PV}+{\bf D}_q\cdot{\bf g}^q_{PV})+2m_1 f^q_P-2d^q_5 f^q_S=0,\nonumber\\
&& \hbar D_q f^q_P-2{\bf \Pi}_q\cdot{\bf g}^q_T-2m_1 f^q_{PV}-2\pi^q_5 f^q_V=0,\nonumber\\
&& \hbar D_q f^q_S-2{\bf \Pi}_q\cdot{\bf g}^q_{T0}+2m_2 f^q_V+2d^q_5 f^q_{PV}=0,\nonumber\\
&& \hbar(D_q{\bf g}^q_{PV}+{\bf D}_qf^q_{PV})-2{\bf \Pi}_q\times{\bf g}^q_V+2m_2{\bf g}^q_T-2\pi^q_5{\bf g}^q_{T0}=0,\nonumber\\
&& \hbar(D_q{\bf g}^q_V+{\bf D}_qf^q_V)-2{\bf \Pi}_q\times{\bf g}^q_{PV}+2m_1{\bf g}^q_{T0}+2d^q_5{\bf g}^q_T=0,\nonumber\\
&& \hbar(D_q{\bf g}^q_{T0}-{\bf D}_q\times{\bf g}^q_T)+2{\bf \Pi}_q f^q_S-2m_1{\bf g}^q_V-2\pi^q_5{\bf g}^q_{PV}=0,\nonumber\\
&& \hbar(D_q{\bf g}^q_T+{\bf D}_q\times{\bf g}^q_{T0})+2{\bf \Pi}_q f^q_P-2m_2{\bf g}^q_{PV}-2d^q_5{\bf g}^q_V=0
\end{eqnarray}
and the constraint equations,
\begin{eqnarray}
\label{constraint}
&& F'_q=\hbar{\bf D}_q\cdot{\bf g}^q_{T0}/2-\Pi_q f^q_S+\pi^q_5 f^q_{PV}+m_1 f^q_V,\nonumber\\
&& P'_q=\hbar{\bf D}_q\cdot{\bf g}^q_T/2-\Pi_q f^q_P+d^q_5 f^q_V+m_2 f^q_{PV},\nonumber\\
&& V'_{0q}={\bf \Pi}_q\cdot{\bf g}^q_V-\Pi_q f^q_V+d^q_5 f^q_P+m_1 f^q_S,\nonumber\\
&& A'_{0q}=-{\bf \Pi}_q\cdot{\bf g}^q_{PV}+\Pi_q f^q_{PV}-\pi^q_5 f^q_S+m_2 f^q_P,\nonumber\\
&& {\bf V}'_q=\hbar{\bf D}_q\times{\bf g}^q_{PV}/2+{\bf \Pi}_q f^q_V-\Pi_q{\bf g}^q_V+\pi^q_5 {\bf g}^q_T+m_2{\bf g}^q_{T0},\nonumber\\
&& {\bf A}'_q=\hbar{\bf D}_q\times{\bf g}^q_V/2-{\bf \Pi}_q f^q_{PV}+\Pi_q{\bf g}^q_{PV}-d^q_5 {\bf g}^q_{T0}+m_1{\bf g}^q_T,\nonumber\\
&& S'_{0iq}{\bf e}_i=-\hbar{\bf D}_qf^q_S/2-{\bf \Pi}_q\times{\bf g}^q_T-\Pi_q{\bf g}^q_{T0}+d^q_5 {\bf g}^q_{PV}+m_2{\bf g}^q_V,\nonumber\\
&& S'_{jkq}\epsilon^{ijk} {\bf e}_i=\hbar{\bf D}_qf^q_P-2{\bf \Pi}_q\times{\bf g}^q_{T0}+2\Pi_q{\bf g}^q_T+2\pi^q_5 {\bf g}^q_V+2m_1{\bf g}^q_{PV}
\end{eqnarray}
for the equal-time spin and isospin components, where $\Gamma'_{iq}(x,{\bf p})=\int dp_0 p_0\Gamma_{iq}(x,p)$ are the first-order energy average of the covariant components $\Gamma_{iq}$, and the equal-time operators are defined as
\begin{eqnarray}
\label{3operator}
&& \Pi_q = iQ_q\hbar\int_{-1/2}^{1/2}dss{\bf E}(t,{\bf x}+is\hbar{\bf \nabla}_p)\cdot{\bf \nabla}_p,\nonumber\\
&& {\bf \Pi}_q ={\bf p} -iQ_q\hbar\int_{-1/2}^{1/2}dss{\bf B}(t,{\bf x}+is\hbar{\bf \nabla}_p)\times{\bf \nabla}_p,\nonumber\\
&& D_q=\partial_t+Q_q\int_{-1/2}^{1/2}ds {\bf E}(t,{\bf x}+is\hbar{\bf \nabla}_p)\cdot{\bf \nabla}_p,\nonumber\\
&& {\bf D}_q = {\bf \nabla}+Q_q\int_{-1/2}^{1/2}ds{\bf B}(t,{\bf x}+is\hbar{\bf \nabla}_p)\times{\bf \nabla}_p,\nonumber\\
&& \pi^q_5 = sgn(Q_q)\sin(\frac{\hbar}{2}{\bf \nabla}\cdot{\bf \nabla}_p)\pi(x),\nonumber\\
&& d^q_5 = sgn(Q_q)\cos(\frac{\hbar}{2}{\bf \nabla}\cdot{\bf \nabla}_p)\pi(x),\nonumber\\
&& m_1 = m_0-\cos(\frac{\hbar}{2}{\bf \nabla}\cdot{\bf \nabla}_p)\sigma(x),\nonumber\\
&& m_2 =\sin(\frac{\hbar}{2}{\bf \nabla}\cdot{\bf \nabla}_p)\sigma(x).
\end{eqnarray}
We have directly used the electromagnetic fields ${\bf E}(x)$ and ${\bf B}(x)$ instead of the tensor $F_{\mu\nu}(x)$.

Explicitly doing the $p_0$-integration on the right-hand side of the relations (\ref{sigmapion}) gives the chiral and pion condensates $\sigma$ and $\pi$ in terms of the equal-time components $f^q_P$ and $f^q_S$,
\begin{eqnarray}
\label{sigmapion2}
&& \sigma(x) = G\int{d^3{\bf p}\over (2\pi)^3}\left[f^u_S(x,{\bf p})+f^d_S(x,{\bf p})\right],\nonumber\\
&& \pi(x) = -G\int{d^3{\bf p}\over (2\pi)^3}\left[f^u_P(x,{\bf p})-f^d_P(x,{\bf p})\right].
\end{eqnarray}

The explicit $\hbar$-dependence of the equal-time transport and constraint equations (\ref{transport}) and (\ref{constraint}) and the operators (\ref{3operator}) helps us to make semi-classical expansion in the equal-time kinetic theory to see clearly the quantum effect order by order. Let us first solve the equations in classical limit $\hbar\to 0$. In this case, quarks behave as quasiparticles with positive and negative energies $p_0=\pm E_p$ and any spin and isospin component includes a positive and a negative energy part, $f^q_i=f_i^{q+} + f_i^{q-}$ and ${\bf g}^q_i={\bf g}_i^{q+} + {\bf g}_i^{q-}$ with $i=0,1,2,3$. From the constraint equations (\ref{constraint}) at the zeroth order in $\hbar$, we can fix the quark energy shell $E_p=\sqrt{m^2+{\bf p}^2}$ with dynamic quark mass $m^2=(m_0-\sigma)^2+\pi^2$ generated by the chiral symmetry breaking and $U_A(1)$ symmetry breaking, and obtain the relations among the classical components,
\begin{eqnarray}
\label{relation}
&& f_{PV}^{q\pm} = \pm{{\bf p}\over E_p}\cdot{\bf g}_{PV}^{q\pm},\nonumber\\
&& f_P^{q\pm} = \pm{sgn(Q_q)\pi\over E_p}f_V^{q\pm},\nonumber\\
&& f_S^{q\pm} = \pm{m_0-\sigma\over E_p}f_V^{q\pm},\nonumber\\
&& {\bf g}_V^{q\pm} =\pm {{\bf p}\over E_p}f_V^{q\pm},\nonumber\\
&& {\bf g}_{T0}^{q\pm} ={{\bf p}\times{\bf g}_{PV}^{q\pm}+sgn(Q_q)\pi{\bf g}_T^{q\pm}\over m_0-\sigma},\nonumber\\
&& {\bf g}_T^{q\pm} = \mp{1\over E_pm^2}\left[E^2_p(m_0-\sigma){\bf g}_{PV}^{q\pm}-(m_0-\sigma)({\bf p}\cdot{\bf g}_{PV}^{q\pm}){\bf p}\mp sgn(Q_q)E_p\pi{\bf p}\times{\bf g}_{PV}^{q\pm}\right].
\end{eqnarray}
These relations greatly simplify the calculation of the equai-time Wigner function: There are only two independent distributions $f_V$ and ${\bf g}_{PV}$, and all the others can be expressed in terms of the two. Note that, the on-shell condition and classical relations come from the constraint equations (\ref{constraint}). If one considers the classical limit of the transport equations (\ref{transport}), one derives only a part of the relations and fails to obtain the on-shell condition. By calculating the physical densities of the system like charge, energy, momentum and angular momentum, we can establish the physical meaning of the components~\cite{birula}. For instance, $f_V$ is the charge density, $f_S$ the mass density, ${\bf g}_{PV}$ the spin density, and ${\bf g}_V$ the current density.

We now integrate the second and third classical relations for $f_P$ and $f_S$ over momentum. By using the expressions (\ref{sigmapion2}) for the condensates, the two order parameters $\pi$ and $\sigma$ are controlled by the gap equations
\begin{eqnarray}
\label{gap}
&& \pi = 0,\\
&& m\left(1+2G\int{d^3{\bf p}\over (2\pi)^3}\sum_{q=u,d}{f_V^{q+}-f_V^{q-}\over E_p}\right)-m_0=0,\nonumber
\end{eqnarray}
where the quark mass is reduced to $m=m_0-\sigma$. The above solution means that, at classical level the chiral symmetry after explicit breaking $U_L(1)\times U_R(1)$ is spontaneously broken but the $U_A(1)$ symmetry is protected. Note that, when we consider a homogeneous system with thermal equilibrium distribution functions $f_V^{q\pm}$, the gap equation for the chiral condensate is exactly the same as the usually used one at finite temperature~\cite{zhuang4}.

From the $\hbar$-structure of the transport equations (\ref{transport}), the linear equations in $\hbar$ govern the dynamical evolution of the classical Wigner function. With the help of the classical relations (\ref{relation}), we obtain the transport equations for the two independent distributions, the number density $f_V^{q\pm}$ and spin density ${\bf g}_{PV}^{q\pm}$,
\begin{eqnarray}
\label{f0g0}
&& \left(D_q \pm {{\bf p}\over E_p}\cdot{\bf D}_q\mp {{\bf \nabla}m^2\cdot{\bf \nabla}_p\over 2E_p}\right)f_V^{q\pm} = 0,\\
&& \left(D_q \pm {{\bf p}\over E_p}\cdot{\bf D}_q\mp {{\bf \nabla}m^2\cdot{\bf \nabla}_p\over 2E_p}\right){\bf g}_{PV}^{q\pm} = {Q_q\over E_p^2}\left[{\bf p}\times \left({\bf E}\times {\bf g}_{PV}^{q\pm}\right)\mp E_p{\bf B}\times{\bf g}_{PV}^{q\pm}\right]-{1\over 2E_p^2m^2}\left(\partial_tm^2{\bf p}\pm E_p{\bf \nabla}m^2\right)\times\left({\bf p}\times{\bf g}_{PV}^{q\pm}\right)\nonumber
\end{eqnarray}
with the reduced operators $D_q=\partial_t+Q_q{\bf E}(x)\cdot{\bf \nabla}_p$ and ${\bf D}_q = {\bf \nabla}+Q_q{\bf B}(x)\times{\bf \nabla}_p$. The first equation for the quark number density is the Boltzmann equation with a mean field force $F=-{\bf \nabla}m^2/(2E_p)$, and the second one describes the quark spin evolution including interaction with the electromagnetic fields on the right hand side. The spin equation is the phase-space version of a generalized Bargmann-Michel-Telegdi (BMT) equation~\cite{bmt}.

The transport equations (\ref{transport}) to the first order in $\hbar$ determine not only the transport equations for the classical distributions $f_V$ and ${\bf g}_0$ but also the first order quantum correction to the pion condensate,
\begin{equation}
\pi^{(1)} = {G\over 2m_0}\int{d^3{\bf p}\over (2\pi)^3}\sum_{q=u,d}Q_q\left[({\bf B}\times{\bf \nabla}_p)\cdot{\bf g}_{PV}^q+{\bf E}\cdot {\bf \nabla}_p f_{PV}^q\right]
\end{equation}
with classical components $f_{PV}$ and ${\bf g}_{PV}$. Taking into account the classical relation between $f_{PV}$ and ${\bf g}_{PV}$, the first order quantum correction comes from the quark spin. While it is difficult to analytically solve the transport equation for the quark spin in general case, we can obtain a solution under the assumption of constant electromagnetic fields,
\begin{equation}
\label{solution}
{\bf g}_{PV}^{q\pm} = {Q_q\over m^2}\left(\mp {\bf B}+{{\bf p}\over E_p}\times{\bf E}\right).
\end{equation}
With this solution we obtain the explicit dependence of the the pion condensate on the electromagnetic fields,
\begin{eqnarray}
\pi^{(1)} &=& -{G\over 2m_0m^2}\left(Q_u^2-Q_d^2\right)\int {d^3{\bf p}\over (2\pi)^3}{2E_p^2+m^2\over E_p^3}{\bf E}\cdot{\bf B}\nonumber\\
&=&-{G\over 4\pi^2 m_0}\left(Q_u^2-Q_d^2\right){\Lambda^3\over m^2\sqrt{\Lambda^2+m^2}}{\bf E}\cdot{\bf B},
\end{eqnarray}
where $\Lambda$ is the momentum cutoff in the model~\cite{zhuang4}. This result is consistent with that obtained in the frame of finite temperature field theory~\cite{cao}. What we want to emphasize here is the quantum effect: Only after considering the quantum correction induced by quark spin, the pion condensate becomes nonzero and $U_A(1)$ symmetry is broken.

The first-order quantum correction to the pion condensate is still in the frame of quasiparticles with mass $m$ determined by the gap equation (\ref{gap}). What is the higher order quantum correction, and what is the full quantum effect on the $U_A(1)$ symmetry breaking? In general case with off-shell effect, all the spin and isospin components $f_i^q$ and ${\bf g}_i^q$ are independent, and we have to solve the full transport equations (\ref{transport}). As a simple example, we consider the following problem. In high energy nuclear collisions, the strong electromagnetic fields exist only in the very beginning and decay very fast~\cite{kharzeev,skokov,deng}. In this case, a natural question is that, if the pion condensate is generated in the beginning via quark spin interaction with the electromagnetic fields, what is its dynamical evolution when the electromagnetic fields disappear? If it vanishes immediately, one cannot explicitly see its effect in the final state distributions. If not, one can directly measure its effect in the final state. This problem is manifested as the following transport equations
\begin{eqnarray}
\label{transport2}
&& \partial_t f_V^q=0,\nonumber\\
&& \partial_t f_{PV}^q +2(m_0-\sigma)f_P^q-2 sgn(Q_q)\pi f_S^q=0,\nonumber\\
&& \partial_t f_P^q -2 {\bf p}\cdot{\bf g}_T^q-2(m_0-\sigma)f_{PV}^q=0,\nonumber\\
&& \partial_t f_S^q -2 {\bf p}\cdot{\bf g}_{T0}^q+2 sgn(Q_q)\pi f_{PV}^q=0,\nonumber\\
&& \partial_t {\bf g}_{PV}^q-2{\bf p}\times{\bf g}_V^q =0,\nonumber\\
&& \partial_t {\bf g}_V^q -2{\bf p}\times{\bf g}_{PV}^q+2(m_0-\sigma){\bf g}_{T0}^q+2 sgn(Q_q)\pi{\bf g}_T^q=0,\nonumber\\
&& \partial_t {\bf g}_{T0}^q +2{\bf p}f_S^q-2(m_0-\sigma){\bf g}_V^q=0,\nonumber\\
&& \partial_t {\bf g}_T^q +2{\bf p} f_P^q-2 sgn(Q_q)\pi{\bf g}_V^q=0
\end{eqnarray}
plus a nontrivial initial condition of nonzero pion condensate. To simplify the transport equations, we have neglected the space dependence of the Wigner function and considered only its time evolution. To be specific, we take the initial condition $-\sigma(t_0)=\pi(t_0)=100$ MeV which then determine the initial components $f_P$ and $f_S$. The other initial components are controlled by the relations (\ref{relation}). The NJL model parameters are chosen as $\Lambda=650$ MeV and $m_0=5$ MeV~\cite{zhuang4}.

\begin{figure}[htb]
{\includegraphics[width=0.40\textwidth]{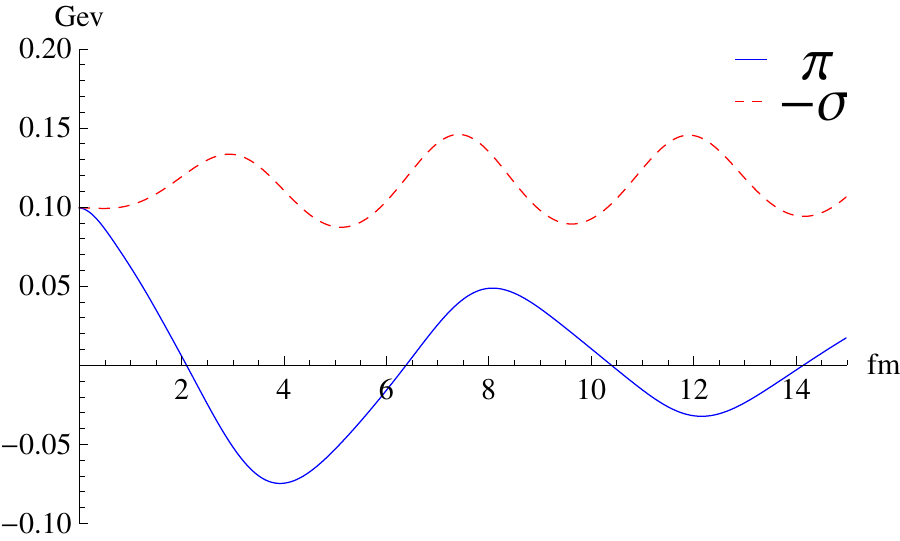}
\caption{ The time evolution of chiral condensate $\sigma(t)$ (dashed line) and pion condensate $\pi(t)$ (solid line). }
\label{fig1}}
\end{figure}

The time evolution of the condensates $\sigma(t)$ and $\pi(t)$ is shown in Fig.\ref{fig1}. In classical limit, the two transport equations (\ref{f0g0}) together with the relations (\ref{relation}) gives trivial solution of the chiral and pion condensates: $\sigma(t)=\sigma(t_0)$ and $\pi(t)=0$ for $t>t_0$. The chiral condensate does not change when the electromagnetic fields disappear at $t_0$, while the pion condensate suddenly drops down at $t_0$ from the initial nonzero value to zero. In quantum case, the multi couplings among all the components shown in (\ref{transport2}) result in strong and long-lived $\sigma$ and $\pi$ oscillations around their classical values. This shows that, when the off-shell effect is taken into account, the multi couplings lead to a significant memory effect on the evolution of the pion condensate: It does not immediately drop down to zero and keep zero during the evolution, but continuously decreases in the early stage and then strongly oscillates around zero.

In summary, we investigated, in the frame of equal-time Wigner function formalism, the quantum dynamical evolution of chiral and $U_A(1)$ symmetry breaking under electromagnetic fields. In a Nambu--Jona-Lasinio model we derived the equal-time transport and constraint equations which determine the quark distributions and the chiral and pion condensates. By taking semi-classical expansion of the equations, we analyzed the quantum effect order by order. In classical limit, while chiral symmetry is spontaneously broken, the $U_A(1)$ symmetry is protected. The first order quantum correction to the $U_A(1)$ symmetry breaking comes from the quark spin interaction with electromagnetic fields. Beyond the quasiparticle approximation, the multi couplings among all the distribution functions lead to a significant memory effect which generates strong and long-lived oscillations for the chiral and pion condensates.

\appendix {\bf Acknowledgement}: The work is supported by the NSFC and MOST grant Nos. 11335005, 11575093 and 2014CB845400.

\end{document}